\documentclass[aps, prd, twocolumn, nofootinbib, showpacs]{revtex4-1}

\usepackage{amsfonts,amsmath,amsthm,amssymb,graphicx,verbatim}

\usepackage{ulem}
\usepackage[usenames]{color}
\definecolor{DarkGreen}{rgb}{0.0,0.5,0.0}
\usepackage{bm}

\def\spose#1{\hbox to 0pt{#1\hss}}

\def\lta{\mathrel{\spose{\lower 3pt\hbox{$\mathchar"218$}}
     \raise 2.0pt\hbox{$\mathchar"13C$}}}
\def\gta{\mathrel{\spose{\lower 3pt\hbox{$\mathchar"218$}}
     \raise 2.0pt\hbox{$\mathchar"13E$}}}
\newcommand{\be}{\begin{equation}}
\newcommand{\en}{\end{equation}}
\newcommand{\bea}{\begin{eqnarray}}
\newcommand{\ena}{\end{eqnarray}}

\newcommand{\dd}{\mbox{d}}

\newcommand{\ie}{\textsl{i.e.~}}

\newcommand{\eg}{\textsl{e.g.~}}
\newcommand{\etal}{\textsl{et al.~}}

 \newcommand{\Ka}{{\cal K}}
 
\newcommand{\GN}{G_{_\mathrm{N}}}

\newcommand{\lP}{\ell_{_\mathrm{Pl}}}

\begin{document}

\preprint{1234}

\title{Stochastic background of relic gravitons in a bouncing quantum cosmological model}

\author{Dennis Bessada${}^{1,3}$\footnote{
        {\tt dbessada@das.inpe.br},\\
        ${}^\dagger\,{}${\tt beatriz@if.ufrj.br},\\${}^\ddagger\,{}${\tt nelsonpn@cbpf.br},\\${}^\S\,{}${\tt oswaldo@das.inpe.br}}, Nelson
        Pinto-Neto${}^{2\ddagger}$,
        Beatriz B. Siffert${}^{2,4\dagger}$ and Oswaldo~D.~Miranda${}^{3\S}$}
\affiliation{
${}^1$UNIFESP - Universidade Federal de S\~ao Paulo - Laborat\'orio de F\'isica Te\'orica e Computa\c c\~ao Cient\'ifica,\\Rua S\~ao Nicolau, 210, 09913-030,
Diadema, SP, Brazil\\
              ${}^2$ICRA - CBPF - Centro Brasileiro de Pesquisas F\'{\i}sicas - Rua Dr. Xavier Sigaud, 150, Urca, 22290-180, Rio de
Janeiro, Brazil\\
              ${}^3$INPE - Instituto Nacional de Pesquisas Espaciais - Av. dos Astronautas, 1758, 12227-010, S\~ao Jos\'e dos Campos, SP,
              Brazil    \\
              ${}^4$Instituto de F\'{\i}sica - Universidade Federal do Rio de Janeiro - Av. Athos da Silveira Ramos 149, 21941-972, Rio de Janeiro, Brazil}

\begin{abstract}

The spectrum and amplitude of the stochastic background of relic gravitons produced
in a bouncing universe is calculated. The matter content of the model consists of
dust and radiation fluids, and the bounce occurs due to quantum cosmological effects
when the universe approaches the classical singularity in the contracting phase. The
resulting amplitude is very small and it cannot be observed by any present and near
future gravitational wave detector. Hence, as in the ekpyrotic model, any observation
of these relic gravitons will rule out this type of quantum cosmological bouncing model.

\end{abstract}


\maketitle

\section{Introduction}
\label{sec:introduction}

Bouncing cosmological models \cite{bouncebn} are being widely investigated because, besides solving the
singularity problem in cosmology by construction, they can also solve the
horizon and flatness puzzles, and lead to an almost scale invariant spectrum
of scalar perturbations if the contracting phase is dominated by dust at large
scales \cite{pert}.

One of the calculations that have been done was the evaluation of the spectral index
of long wavelength tensor perturbations, $n_T$, and in most of bouncing models they
were found to be also scale invariant. With respect to their amplitudes, specific models
must be worked out. For instance, in the cyclic ekpyrotic scenario, they were evaluated
and it was shown that the amplitudes are too small to be detected by present gravitational
wave detectors \cite{cyclic}.

In this paper we will calculate the spectrum and amplitude of relic gravitons in a different bouncing
cosmological model. It consists of a Friedmann-Lema\^{\i}tre-Robertson-Walker (FLRW) universe
filled by dust and radiation, which is contracting classically. As it approaches the singularity,
quantum cosmological effects on the background, here described through the Wheeler-DeWitt equation
interpreted along the lines of the Bohm-de Broglie quantum theory, avoids the singularity and ejects
the universe to the expanding phase we are now experiencing.
Note that our bouncing model is very conservative: there is nothing else than dust and radiation, we are
working with a $3+1$ dimensional space-time, and we are performing a canonical quantization of the second
order perturbed (with respect to the FLRW background) Einstein-Hilbert action of general relativity,
interpreted along the lines of a quantum theory appropriate to quantum cosmology, namely, one which
does not need any external agent to the quantum system to give a meaning to the quantum calculations.

Evolution of quantum perturbations (scalar, vector
and tensor) on these quantum backgrounds can be described by simple equations, as it was demonstrated
in Refs.~\cite{Peter:2006id,nelson}. With these equations, we were able to calculate the spectrum (analytically
and numerically) and amplitude (numerically) of this stochastic background of relic gravitons. Although,
the spectrum and amplitude differ considerably from the cyclic ekpyrotic scenario, the main conclusion
remains the same, namely, that the amplitudes are too small to be detected by any present and near future
gravitational wave detectors.

The paper is divided as follows: in the next section we review the main aspects of the quantum
cosmological bouncing model on which the relic gravitons evolves, and we obtain the dynamical equations
that the tensor perturbations which describe these relic gravitons must obey. In section III we derive
the expression for the critical fraction of the relic gravitons energy density from the tensor
perturbations described in section II. In section IV we calculate the spectrum and amplitude
of this energy density and the graviton strain either analytically (for the spectrum) and numerically.
We end up with the conclusions.

\section{The stochastic background of relic gravitons in a bouncing quantum cosmological model}
\label{sec:qcosmomodel}

We start reviewing the key ideas concerning tensor perturbations
in a perfect fluid quantum cosmological model in the Bohm-de
Broglie interpretation as put forward in Refs.
\cite{Peter:2005hm,Peter:2006id}.

\subsection{Hamiltonian formalism for tensor perturbations}
\label{subsec:hamfor}

We start with an Einstein-Hilbert action coupled to a perfect
fluid described by the Schutz formalism \cite{Schutz}:
\begin{equation}
\label{action} S = S_{\mathrm{GR}} + S_\mathrm{fluid} =
-\frac{1}{6\lP^2} \int \sqrt{-g} R \dd^4 x + \int \sqrt{-g} P\dd^4 x,
\end{equation}
where $\lP=(8\pi\GN/3)^{1/2}$ is the Planck length in natural
units ($\hbar=c=1$), $P$ is the perfect fluid pressure whose
density $\rho$ is given by the equation of state $P=\omega\rho$,
with $\omega=\mathrm{const}$.

Writing down the action of a fluid as proportional to the pressure has also been
proposed by other authors \cite{fluids}. All these approaches are generally covariant, but they yield a preferred time
direction, the one connected with the surfaces of constant potential of the velocity field of the fluid.

Such very simple fluids can also be completely characterized by
the k-essence lagrangian $(\partial_\mu \phi \partial^\mu \phi)^{(w+1)/(2w)}$,
where $\phi$ is a scalar field and $w$ is a constant. This scalar field has equation of state $p=w\rho$.
It has no potential and a non-trivial kinetic term.
The matter Hamiltonian which we will exhibit below can be easily obtained from this Lagrangian after one
performs a simple canonical transformation.

This fluid description is suitable to describe the primordial
universe, when radiation dominates and all particles become relativistic, if we make the choice $w=1/3$.
In what folows, we will directly quantize this perfect fluid Lagrangian. This can be justified in physical grounds
because it was implementes while studying superfluids \cite{landau}, and the quantum model based on this
procedure turned out to be quite accurate in describing many properties of such
physical systems.

The metric
$\mbox{\boldmath$g$\unboldmath}$ in Eq.~(\ref{action}) is
decomposed into a background
Friedman-Lema\^{\i}tre-Robertson-Walker (FLRW) metric and into a
first-order tensor perturbation $w_{ij}$, and is given in the
Arnowitt-Deser-Misner (ADM) formalism by
\begin{equation}
\dd s^{2}=N^2\left(\tau\right)\dd\tau^2 -a_{\rm
  phys}^2\left(\tau\right) \left(\gamma_{ij}+w_{ij}\right)\dd x^{i}\dd
x^{j}. \label{adm2}
\end{equation}
The background metric $\gamma_{ij}$ is related to the spacelike
hypersurfaces with constant curvature $\Ka$ ($\Ka=0,\pm 1$ for
flat, open and closed space respectively), and lowers and raises
the indices of the tensor perturbation $w_{ij}$, which is
transverse and traceless (\ie, $w^{ij}_{\ \ |j}=0$ and $w^i_{\
i}=0$, where the bar indicates a covariant derivative with respect
to $\gamma$). $N(\tau)$ is the lapse function and defines the
gauge, fixed once and for all.

The second-order Hamiltonian for the gravitational model described
by the action (\ref{action}) and metric (\ref{adm2}) can be
written as (see Ref. \cite{Peter:2005hm} for further details)
\begin{widetext}
\begin{eqnarray}H&\equiv&NH_0\nonumber \\
&=&N\left\{-\frac{P_a^{2}}{4a}-\Ka a+ \frac{P_{_T}}{a^{3\omega}}
\left(1+\displaystyle\frac{\omega}{4} \int \dd^{3} x
\gamma^{1/2}\, w_{ij} w^{ij}\right) + \frac{5P_a^2}{48a} \int
\dd^3 x \gamma^{1/2}\, w_{ij}w^{ij} \right. \nonumber\\&& + \left.
\int \dd^3 x \left[ \frac{6\Pi_{ij} {{\Pi}}\null^{ij}}{a^{3}
\gamma^{1/2}} + 2\frac{P_a w_{ij}{\Pi}^{ij}}{a^2} + \gamma^{1/2}a
\left( \frac{w^{ij|k} w_{ij|k} }{24} + \frac{\Ka}{6} w_{ij} w^{ij}
\right)\right] \right\}, \label{h1}
\end{eqnarray}
\end{widetext}
where the quantities $P_a, {\Pi}^{ij}, P_{_T}$ are the momenta
canonically conjugate to the scale factor, the tensor
perturbations, and to the fluid degree of freedom, respectively.
These quantities have been redefined in order to be dimensionless.
For instance, the physical scale factor $a_{\rm phys}$ can be
obtained from the dimensionless $a$ present in~(\ref{h1}) through
$a_{\rm phys}=\lP a/\sqrt{V}$, where $V$ is the comoving volume of
the background spacelike hypersurfaces.  This Hamiltonian, which
is zero due to the constraint $H_0\approx0$, yields the correct
Einstein equations both at zeroth and first order in the
perturbations, as can be checked explicitly. In order to obtain
its expression, no assumption has been made about the background
dynamics, just Legendre and canonical transformations have been
performed.

The fact that the momentum $P_{_T}$ appears linearly in the Hamiltonian
suggests to consider its canonical position $T$ as a time variable. Indeed,
from the canonical transformation used to arrive at this expression \cite{fluids},
one has that $\phi = T$, ($L_m = (\partial_\mu \phi \partial^\mu \phi)^{(w+1)/(2w)}$)
that is, it is just the potential of the fluid velocity, which characterizes
a prefered foliation of spacetime. This time variable always increases,
as it can be checked in the contracting and expanding classical solutions, and in the
Bohmian bounce solution we present below.

In the quantum regime, this Hamiltonian can be substantially
simplified through the implementation of the quantum canonical
transformation generated by
\begin{equation}
\label{u2} U=\exp(iG_{\rm q})
\equiv\exp\left(\frac{i}{12}\hat{\beta}_a \hat{Q}\right),
\end{equation}
where $\hat{\beta}_a \equiv \frac{1}{2} \left(\hat{P}_a \hat{a} +
\hat{a}\hat{P}_a\right)$ and $\hat{Q} \equiv \int \dd^3 x\,
\gamma^{1/2} \hat{w}_{ij} \hat{w}^{ij}$ are the self-adjoint
operators associated with the corresponding classical variables,
yielding, for a particular factor ordering of~(\ref{h1}) (see
Section 3 in Ref.~\cite{Peter:2005hm} for further details), the
following quantum Hamiltonian
\begin{eqnarray}
\hat{H}_0 & = &\left[ -\frac{1}{4\hat{a}}\hat{P}_a^2 -\Ka\hat{a}+
\frac{\hat{P}_{_T}}{\hat{a}^{3\omega}}+\int \dd^3 x \left(
6\frac{\hat\Pi^{ij}\hat{\Pi}_{ij}}{\gamma^{1/2} a^3}
\right.\right.\nonumber \\ & + & \left.\left.\frac{1}{24}
\gamma^{1/2} a \hat{w}_{ij|k}\hat{w}^{ij|k}+ \frac{1}{12}
\gamma^{1/2} \Ka \hat{w}_{ij} \hat{w}^{ij} a \right)\right].
\label{h200}
\end{eqnarray}

Performing an inverse Legendre transform on the second-order piece
of the Hamiltonian (\ref{h200}), and restoring the constant $\lP$,
we get the following Lagrangian density
\begin{eqnarray}
\label{lagtens}
\mathcal{L}^{(2)}=\frac{1}{24\lP^2}\sqrt{-{}^{(0)}g}\left[{}^{(0)}g^{{\alpha\beta}}{w^i}_{j|\alpha}{w_i}^{j|\beta}-2\Ka
{w^i}_j{w^j}_i\right],
\end{eqnarray}
where ${}^{(0)}g_{{\alpha\beta}}$ is the background piece of the
full metric (\ref{adm2}). Note that Lagrangian (\ref{lagtens})
coincides with the one derived in \cite{MFB} for classical
backgrounds.

\subsection{Quantum evolution of the background and perturbation variables}
\label{subsec:backbohm}

The quantization procedure of the background and tensor
perturbations can be implemented by imposing
${\hat{H}}_0\Psi(a,w_{ij})=0$. The Wheeler-DeWitt equation in this
case reads
\begin{widetext}
\begin{eqnarray}
i\frac{\partial\Psi}{\partial T} & = & \hat{H}_\mathrm{red} \Psi
\nonumber \\
&\equiv&\left\{ \frac{a^{3\omega-1}}{4} \frac{\partial^2}{\partial
a^{2}} - \Ka a^{3\omega+1}+ \int \dd^3 x \left[ - 6
\frac{a^{3(\omega-1)}}{\gamma^{1/2}} \frac{\delta^2}{\delta w_{ij}
\delta w^{ij}} + a^{3\omega+1} \left(\gamma^{1/2} \frac{w_{ij|k}
w^{ij|k}}{24} + \Ka\frac{w_{ij}
w^{ij}}{12}\right)\right]\right\}\Psi, \label{es2}
\end{eqnarray}
\end{widetext}
where we have chosen $T$ as the time variable, which is equivalent
to impose the time gauge $N=a^{3\omega}$. Next, using the Bohm-de
Broglie interpretation of quantum mechanics \cite{onto},
making the separation ansatz for the wave functional
$\Psi[a,w_{ij},T]=\varphi(a,T)\psi[a,w_{ij},T]$, and following
the reasoning of Ref.~\cite{felipe}\footnote{We are assuming that there is a disentangled
zeroth order term in the total wave function.
It is a reasonable physical assumption as long as the semi-classical theory of
cosmological perturbations, where the background behaves classically and perturbations are quantum,
seems to describe quite well our real universe. This would be impossible if the wave function were
completely entangled. Note that there is no definite theory of initial conditions for the wave function
of the universe, hence one must rely in such post-factum assumptions, one, namely, that our universe
has a classical limit.}, we can show that Eq.~(\ref{es2}) can be split into
two, namely
\begin{equation}
i\frac{\partial\varphi}{\partial T} =\frac{a^{3\omega-1}}{4}
\frac{\partial^2\varphi}{\partial a^2} - \Ka a^{3\omega+1}\varphi,
\label{es20}
\end{equation}
and
\begin{eqnarray}
i\frac{\partial\psi}{\partial T}&=& \int \dd^3 x
\left[-6\frac{a^{3(\omega-1)}}{\gamma^{1/2}}
\frac{\delta^2}{\delta w_{ij} \delta w^{ij}} \right. \nonumber \\
&+& \left. a^{3\omega+1} \left(\gamma^{1/2} \frac{w_{ij|k}
w^{ij|k}}{24} + \Ka\frac{w_{ij} w^{ij}}{12}\right)\right]\psi.
\label{es22}
\end{eqnarray}

Using the Bohm-de Broglie interpretation, Eq.~(\ref{es20}) can now be solved
as in Refs.~\cite{pinto,pinto2,fabris}, yielding a Bohmian quantum
trajectory $a(T)$, which in turn can be used to simplify Eq.~(\ref{es22})
\cite{Peter:2005hm}. Indeed, as one can understand $a(T)$ as a prescribed function
of time, which implies that $\psi(a,w_{ij} , T)=\psi(a(T),w_{ij} , T)=\psi '(w_{ij} , T)$,
one can perform the time dependent unitary transformation

\begin{widetext}

\begin{equation}
\label{transformacao unitaria}
U = \exp\biggl\{ i
\biggl[ \int d^3 x \gamma^{1/2} \frac{a^{\prime} w_{ij} w^{ij}}{2a}
\biggr] \biggr\} \exp\biggl\{ i \biggl[ \int d^3 x \biggl(
\frac{w_{ij}\Pi ^{ij} + \Pi ^{ij} w_{ij}}{2} \biggr)
\ln\biggl( \frac{\sqrt{12}}{a} \biggr) \biggr]\biggr\},
\end{equation}
yielding the following simple form for the Schr\"odinger equation for the perturbations:

\begin{equation}
\label{simplest}
i\frac{\partial\varphi(w,\eta)}{\partial \eta}= \int d^3 x
\left\{-\frac{1}{2\gamma^{1/2}} \frac{\delta^2}{{\delta w}^2} +
\gamma^{1/2}\left[\frac{1}{2}w_{k} w^{k}
-\frac{a^{\prime\prime}}{2a}w^2\right] \right\} \varphi(w,\eta).
\end{equation}
A transformation to conformal time
$\eta$, $a^{3\omega-1}\dd T = \dd \eta$, was also done, and
a prime $'$ denotes the derivative with respect to $\eta$ (see again Ref.~\cite{Peter:2005hm} for
details).

This Schr\"odinger equation is identical to the one used in semi-classical
gravity for linear tensor perturbations, but here it was obtained without ever
using the background Einstein's equations. Hence it can be used when the background
is also quantized. It is important to stress
that the function $a(\eta)$ which is present in Eq.~(\ref{simplest}) is
no longer the classical solution for the scale factor, but the
quantum Bohmian solution. This fact leads to some different consequences with
respect to the usual semi-classical approach, as we will see.

In the Heisenberg representation, the equations for the operator
${\hat{w}}_{ij}$ read

%Eq. (\ref{es22})turns into \cite{Peter:2006id}

\begin{equation}
\label{tenspereq}{\hat{w}}_{ij}''+2\frac{a'}{a}{\hat{w}}_{ij}' -
{\hat{w}}_{ij|k}^{|k}+ 2\Ka {\hat{w}}_{ij}
 = 0,
\end{equation}
which corresponds to the usual equation for quantum tensor perturbations
in classical backgrounds \cite{MFB}.

We can proceed with the usual analysis, but now taking the
quantum Bohmian solution $a(\eta)$ coming from Eq.~(\ref{es20})
as the new pump field. In order to obtain these background
quantum solutions, we follow the procedure discussed in Ref.
\cite{Peter:2006id} which, for flat spatial sections $\left(\Ka =
0\right)$, yields the following guidance relation
\begin{equation}
\label{guidance} \frac{da}{dT}=-\frac{a^{(3\omega-1)}}{2}
\frac{\partial S}{\partial a},
\end{equation}
in accordance with the classical relations $da/dT=\{a,H\}=
-\frac{1}{2}a^{(3\omega-1)}P_a$ and $P_a=\partial S/\partial a$.
For the wave function

\begin{equation}\label{psi1t}
\Psi(a,T)=\left[\frac{8 T_b}{\pi\left(T^2+T_b^2\right)}
\right]^{1/4}
\exp\biggl[\frac{-4T_ba^{3(1-\omega)}}{9(T^2+T_b^2)(1-\omega)^2}\biggr]
\exp\left\{-i\left[\frac{4Ta^{3(1-\omega)}}{9(T^2+T_b^2)(1-\omega)^2}
+\frac{1}{2}\arctan\biggl(\frac{T_b}{T}\biggr)-\frac{\pi}{4}\right]\right\},
\end{equation}
which comes from the initial normalized gaussian at $T=0$

\begin{equation}
\label{initial}
\Psi^{(\mathrm{init})}(\chi)=\biggl(\frac{8}{T_b\pi}\biggr)^{1/4}
\exp\left(-\frac{\chi^2}{T_b}\right) ,
\end{equation}
\end{widetext}
where $T_b$ is an arbitrary constant,
the Bohmian quantum trajectory for the scale factor
is given by
\begin{equation}
\label{at} a(T) = a_b
\left[1+\left(\frac{T}{T_b}\right)^2\right]^\frac{1}{3(1-\omega)},
\end{equation}
where $a_b$ is the minimum value for the scale factor at the
bounce $T=0$. Note that this solution has no singularities and
tends to the classical solution when $T\rightarrow\pm\infty$.
The quantity $T_b$, which is the width of the initial gaussian (15),
yields the curvature scale at the bounce,
$L_{\rm bounce}\equiv T_b a_b^{3w}$. There is nothing,
at this stage of quantum cosmology, which may constrain its value: we have not
yet a theory of initial conditions. It is a free parameter.
However, physical considerations allow us to say that it must be greater than
the Planck length but smaller than the curvature scale when nucleosynthesis
takes place, in order to not spoil its results.

The tensor perturbations are quantum-mechanical operators, hence it is convenient to expand them into Fourier
modes and subject them to quantization rules:
\begin{eqnarray}
\label{expw} {\hat{w}}_{ij} \left(x\right) & = & \sqrt{6}\lP
\sum_{\lambda = +,\times}\int \frac{d^3
k}{\left(2\pi\right)^{3/2}}
\varepsilon^{(\lambda)}_{ij}\left[w^{(\lambda)}_k\left(\eta\right)
e^{-i\mathbf{k}\cdot\mathbf{x}}\hat{a}_\mathbf{k}^{(\lambda)}\right.\nonumber
\\ & + & \left.w^{\ast(\lambda)}_k\left(\eta\right)
e^{i\mathbf{k}\cdot\mathbf{x}}\hat{a}_\mathbf{k}^{(\lambda)\dagger}
\right],
\end{eqnarray}
where $x=\left(\eta,\mathbf{x}\right)$,
$\varepsilon^{(\lambda)}_{ij}=\varepsilon^{(\lambda)}_{ij}\left(\hat{\mathbf{k}}\right)$
is the polarization tensor for the two graviton polarization
states $+$ and $\times$ labeled by $\lambda$, and satisfies
\begin{equation}
\label{polten}\varepsilon^{(\lambda)ij}\varepsilon^{(\lambda')}_{ij}=2\delta_{\lambda\lambda'}.
\end{equation}
Also, $w^{(\lambda)}_k\left(\eta\right)$ are mode functions, and
$\hat{a}_\mathbf{k}^{(\lambda)\dagger}$,
$\hat{a}_\mathbf{k}^{(\lambda)}$ are creation and annihilation
operators, respectively. Such operators satisfy the equal-time
commutation relations
\begin{eqnarray}
\label{comma1}
&&\left[\hat{a}_\mathbf{k}^{(\lambda)},\hat{a}_\mathbf{k'}^{(\lambda')\dagger}\right]
= \delta_{\lambda\lambda'}
\delta^{(3)}\left(\mathbf{k}-\mathbf{k'}\right),\\
&&\left[\hat{a}_\mathbf{k}^{(\lambda)},\hat{a}_\mathbf{k'}^{(\lambda')}\right]
=
\left[\hat{a}_\mathbf{k}^{(\lambda)\dagger},\hat{a}_\mathbf{k'}^{(\lambda')\dagger}\right]
= 0,\label{comma2}
\end{eqnarray}
and the quantum vacuum is defined by
\begin{equation}
\label{defvac} \hat{a}_\mathbf{k}^{(\lambda)}| 0 \rangle = 0.
\end{equation}

This vacuum initial condition was chosen because the universe in the contracting phase was very
big, rarefied and almost flat, where inhomogeneities are supposed to be
wiped out through dissipation, see Ref.~\cite{nelson} on that. This is a general
assumption in bouncing models.

Next, inserting the Fourier expansion into Eq. (\ref{tenspereq}),
we get the mode equation
\begin{equation}
\label{modeeq} w^{(\lambda)\prime\prime}_k +
2\frac{a'}{a}w^{(\lambda)\prime}_k + \left(k^2 +2\Ka\right)
w^{(\lambda)}_k = 0.
\end{equation}
Introducing the canonical amplitude $v^{(\lambda)}_k$ as
\begin{equation}
\label{defvk} v^{(\lambda)}_k\equiv aw^{(\lambda)}_k,
\end{equation}
mode equation (\ref{modeeq}) becomes
\begin{equation}
\label{modeequation} v_k^{(\lambda)''}+\left( k^2 + 2\Ka
-\frac{a''}{a} \right)v^{(\lambda)}_k =0,
\end{equation}
for each graviton polarization state. In the present work, we will only be concerned with the case $ \Ka=0$, i.e., a Universe with flat spatial sections.

\section{Stochastic Background of Relic Gravitons}
\label{sec:stochbac}

In the $\Lambda$CDM model, quantum fluctuations arising during
inflation lead to a nearly scale-invariant spectrum of density
(scalar) \cite{inflation2} and gravitational waves (tensor) perturbations
\cite{gravitons}. As in the case of CMB, we also expect relic
gravitons generated in this early epoch to form a background --
the stochastic background of relic gravitons (SBRG) -- to be
hopefully detected by the high-sensitivity gravitational waves (GW)
detectors. The physical observable to be measured by such GW
detectors is the critical fraction of the relic gravitons energy
density $\rho_{GW}$ given by (see
\cite{Giovannini:2009kg,Maggiore:1999vm} and references therein)
\begin{equation}
\label{defOmega}
\Omega_{GW}\left(\eta\right)\equiv\frac{\rho_{GW}\left(\eta\right)}{\rho_{c}},
\end{equation}
where $\rho_{c}=\left(H/\lP\right)^2$ is the critical energy
density. The energy density $\rho_{GW}$ carried by the relic
gravitons is simply the 0-0 component of the stress
energy-momentum tensor of the tensor perturbations, that is
$\rho_{GW}={T^0}_{0}$, where
\begin{equation}
\label{defemtgw} T_{\alpha\beta}\equiv
-\frac{2}{\sqrt{-{}^{(0)}g}}\frac{\partial
\mathcal{L}^{(2)}}{\partial {}^{(0)}g^{{\alpha\beta}}},
\end{equation}\\
and $\mathcal{L}^{(2)}$ is the Lagrangian (\ref{lagtens}) for the
tensor perturbations (recall that Lagrangian (\ref{lagtens}) holds
either in classical or quantum backgrounds). Considering only
spatially-flat models $\left(\Ka = 0\right)$, the energy density
of the relic gravitons can be derived by inserting (\ref{lagtens})
into (\ref{defemtgw}) and taking the 0-0 component, which yields
the classical expression
\begin{eqnarray}
\label{rhogw}
\rho_{GW}=\frac{1}{12\lP^2}\left[\frac{1}{a^2}{w^{ij}}'{w_{ij}}'-\frac{1}{2}
w^{ij,\mu}{w_{ij,\mu}}\right].
\end{eqnarray}

Since tensor fluctuations generated during inflation are of
quantum nature, the energy density of the relic gravitons becomes
the expected value of the operator ${{\hat{T}^0}}{}_{0}$ in the
vacuum state defined in (\ref{defvac}):
\begin{eqnarray}
\label{defquantrho} \rho_{GW}\equiv\langle
0|{{\hat{T}^0}}{}_{0}|0\rangle.
\end{eqnarray}
Substituting the Fourier expansion (\ref{expw}) into
(\ref{rhogw}), and the result into (\ref{defquantrho}), we find,
using the commutation relations
(\ref{comma1},\ref{comma2},\ref{defvac}), that
\begin{eqnarray}
\label{quantrho} \rho_{GW}\left(\eta\right) & =
&\frac{1}{a^2}\sum_{\lambda}\int d\ln k\frac{k^3}{4\pi^2} ~
\left[\left|w^{(\lambda)'}_k\left(\eta\right)\right|^2\right.\nonumber
\\ &+& \left. k^2\left|w^{(\lambda)}_k\left(\eta\right)\right|^2\right].
\end{eqnarray}

In terms of the canonical field $v^{(\lambda)}_k$ introduced in
(\ref{defvk}), expression (\ref{quantrho}) turns into
\begin{eqnarray}
\label{quantrhov} \rho_{GW}\left(\eta\right)  = \int d\ln k ~
\rho_{GW}\left(k,\eta\right),
\end{eqnarray}
where we have introduced the energy density per mode
\begin{eqnarray}
\label{defquantrhomode}
\rho_{GW}\left(k,\eta\right)&\equiv&\frac{d\rho_{GW}\left(\eta\right)}{d\ln
k}\nonumber \\ & = &
\frac{k^3}{4\pi^2a^4}\sum_{\lambda}\left\{\left|v^{(\lambda)'}_k\right|^2-\mathcal{H}\left[v^{(\lambda)}_k
v^{(\lambda)'\ast}_k\right.\right. \nonumber \\
&+& \left. \left. v^{(\lambda)\ast}_k v^{(\lambda)'}_k\right]+
\left(k^2 +
\mathcal{H}^2\right)\left|v^{(\lambda)}_k\right|^2\right\},
\end{eqnarray}
where $\mathcal{H}=a'/a$ is the Hubble parameter in conformal time
coordinates.

From (\ref{defquantrhomode}) the energy density parameter
(\ref{defOmega}) per mode is given by
\begin{widetext}
\begin{eqnarray}
\label{defomegamode} \Omega_{GW}\left(k,\eta\right) =
\frac{\lP^2}{2\pi^2a^2}\frac{k^3}{\mathcal{H}^2}\sum_{\lambda}\left\{\left|v^{(\lambda)'}_k\right|^2-\mathcal{H}\left[v^{(\lambda)}_k
v^{(\lambda)'\ast}_k + v^{(\lambda)\ast}_k
v^{(\lambda)'}_k\right]+ \left(k^2 +
\mathcal{H}^2\right)\left|v^{(\lambda)}_k\right|^2\right\}.
\end{eqnarray}
\end{widetext}

By solving the mode equation (\ref{modeequation}) and inserting
the result into (\ref{defomegamode}), we promptly determine the
energy density parameter per logarithmic frequency interval $\nu =
k/(2\pi a)$, which is the physical quantity to be confronted with
future observations.

\section{Stochastic Background of Relic Gravitons in a Dust-Radiation Bouncing Universe}
\label{sec:stochbacq}

\subsection{Scale Factor}
The scale factor for the radiation dominated bounce will be given by Eq. (\ref{at}), with $w=1/3$. In this case, we have $dT=d \eta$ so, we find

\begin{equation}
\label{abounce}
a_{bounce}(\eta)=a_b\sqrt{1+\left(\frac{\eta}{\eta_b}\right)^2},
\end{equation}
where $\eta_b$ is a free parameter that determines the duration of the bouncing phase.

The scale factor before and after the bounce is the usual classical solution for a Universe filled with
radiation and dust \cite{MukhanovBook},

\begin{equation}
\label{aMukhanov}
a_{rad}(\eta)=a_{eq} \left[ \left( \frac{\eta}{\eta_*} \right)^2 \mp 2 \frac{\eta}{\eta_*} \right],
\end{equation}
where the minus and plus signs refers to the epochs before and after the bounce, respectively, $a_{eq}$ is the value of the scale factor at matter-radiation equality, and

\begin{eqnarray}
\eta_*= 2R_H\frac{\sqrt{2+z_{eq}}}{1+z_{eq}},
\end{eqnarray}
where $R_H=1/(a_0H_0)$ is the co-moving Hubble radius, $a_0$ and $H_0$ are the present values of the
scale factor and Hubble radius, respectively, and $z_{eq} \approx 3\times 10^3$ is the value of the redshift at matter-radiation equality. As the bounce should occur much before the matter-radiation
equality, then $\eta_b << \eta_*$.

The above expressions for $a_{bounce}$ and $a_{rad}$ can be condensed into

\begin{equation}
\label{ScaleFactor}
a(\eta)=a_{eq} \left[ \left(\frac{\eta}{\eta_*} \right)^2 + 2 \frac{\eta_b}{\eta_*} \sqrt{1+\left(\frac{\eta}{\eta_b}\right)^2} \right]\,\,,
\end{equation}
which promptly recovers Eqs. (\ref{abounce}) and (\ref{aMukhanov}) in the limits $\eta<<\eta_*$ and $\eta >> \eta_b$, respectively, and assuming that $a_b = 2 a_{eq} \eta_b / \eta_*$.

With this expression for the scale factor, we are ready to solve the mode equation, Eq. (\ref{modeequation}).

\subsection{Mode Equation} \label{sec:lambconst}
\label{subsec:modeeq}

For numerical purposes, we prefer to work with the normalized variables

\begin{equation}
\bar{\eta} \equiv \frac{\eta}{2R_H}\,\,\,\,\, {\rm and} \,\,\,\,\, n \equiv 2R_H k.
\end{equation}

In terms of these variables, the mode equation for a flat Universe becomes

\begin{equation}
\label{FinalModeEq}
\frac{d^2v_n}{d\bar{\eta}^2}+\left[n^2-\frac{d^2 a}{d\bar{\eta}^2} \frac{1}{a} \right] v_n=0,
\end{equation}
where we are treating separately each possible graviton polarization state, so we have dropped the superscript $(\lambda)$.

The scale factor given in Eq. (\ref{ScaleFactor}) becomes

\begin{equation}
\label{FinalScaleF}
a(\bar{\eta})=a_{eq} \left[ \left(\frac{\bar{\eta}}{\bar{\eta}_*} \right)^2 + 2 \frac{\bar{\eta}_b}{\bar{\eta}_*} \sqrt{1+\left(\frac{\bar{\eta}}{\bar{\eta}_b}\right)^2} \right]\,\,,
\end{equation}
where $\bar{\eta}_b=\eta_b/2R_H$ is a free parameter and
\begin{eqnarray}
\bar{\eta}_*=  \frac{\eta_*}{2R_H}=\frac{\sqrt{2+z_{eq}}}{1+z_{eq}} \approx 0.018.
\end{eqnarray}

In the original set of variables $\eta$ and $k$, the initial conditions to solve the mode equation are the vacuum conditions to be set at $\eta_{ini}=-\infty$:

\begin{equation}
v_k(\eta_{ini})=\frac{e^{-ik\eta_{ini}}}{\sqrt{2k}},
\end{equation}
\begin{equation}
\frac{dv_k}{d\eta}\bigg|_{\eta=\eta_{ini}}=-i \sqrt{\frac{k}{2}}e^{-ik\eta_{ini}}.
\end{equation}

Using the new variables $\bar{\eta}$ and $n$, the initial conditions to solve Eq.~(\ref{FinalModeEq}) for one of the two polarization modes are

\begin{equation}
v_n(\bar{\eta}_{ini})=\sqrt{\frac{R_H}{n}}\,e^{-in\bar{\eta}_{ini}},
\end{equation}
\begin{equation}
\frac{dv_n}{d\bar{\eta}}\bigg|_{\bar{\eta}=\bar{\eta}_{ini}}=-i \sqrt{nR_H}\,e^{-in\bar{\eta}_{ini}}.
\end{equation}

We have checked that we can, in this case, numerically consider $-\infty$ as any value of $\bar{\eta}_{ini}<-3$.

The present value of $\bar{\eta}$, $\bar{\eta}_0$, is obtained by solving Eq. (\ref{FinalScaleF}) for $\bar{\eta}=\bar{\eta}_0$, which results in

\begin{equation}
\bar{\eta}_0=-\frac{\sqrt{2-z_{eq}}+2+z_{eq}}{1+z_{eq}} \approx 0.982.
\end{equation}

We can constrain the free parameter $\bar{\eta}_b$ by demanding that the end of the bouncing phase happens between the Planck time ($z \approx 10^{32}$ and the beginning of nucleosynthesis ($z \approx 10^{10}$), which results in

\begin{equation}
10^{-31} \lesssim \bar{\eta}_b \lesssim 10^{-9}.
\end{equation}

We have solved Eq. (\ref{FinalModeEq}) from $\bar{\eta}_{ini}$ to $\bar{\eta}_0$ for four different values of $\bar{\eta}_b$ in this range: $10^{-30}$, $10^{-24}$, $10^{-18}$ and $10^{-12}$.

In terms of the new variables $\bar{\eta}$ and $n$, the expression for the density parameter in Eq. (\ref{defomegamode}), for one of the two possible polarizations, can be rewritten as

\begin{eqnarray}
 \Omega_{GW}\left(n,\bar{\eta}\right) =
\frac{\lP^2}{2\pi^2a^2}\frac{n^3}{32R_H^5\mathcal{H}^2} \left \{ \left|\frac{dv_n}{d\bar{\eta}}\right|^2- 2R_H\mathcal{H} \times {}\nonumber \right.\\\left.
\times \left[v_n \frac{dv^{\ast}_n}{d\bar{\eta}} + v^{\ast}_n  \frac{dv_n}{d\bar{\eta}}\right]+ \left(n^2 + 4R_H^2\mathcal{H}^2\right)\left|v_n\right|^2 \right \}.{}
\end{eqnarray}

For $\bar{\eta}=\bar{\eta}_0$, this expression reduces to

\begin{eqnarray}
\Omega_{GW}(n,\bar{\eta}_0) = \frac{l_p^2}{64\,\pi^2}\, n^3a_0H_0^3 \left\{\bigg| \frac{dv_n}{d\bar{\eta}} \bigg|^2 - 2 \left[v_n\frac{dv^{\ast}_n}{d\bar{\eta}} + \right. \right.{} \nonumber\\ \left. \left.
{}+v^{\ast}_n\frac{dv_n}{d\bar{\eta}} \right] + (n^2+4)|v_n|^2 \right\}   \bigg|_{\bar{\eta}=\bar{\eta}_0}.
\end{eqnarray}

\subsection{Analytical Results} \label{sec:ana}

One can anticipate the spectral behavior of the above quantities through
some analytical reasonings. Note first that the spectral index of tensor perturbations at the moment they get
smaller than the curvature scale
in the expanding phase of the above quantum bouncing models was calculated in
Ref.~\cite{Peter:2006id}, and it reads

\begin{equation}
\label{nt}
n_T = \frac{12 w}{1+3w},
\end{equation}
where $w$ is the equation of state parameter of the fluid which dominates the cosmic evolution
when the mode is getting bigger than the curvature scale of the Universe in the contracting phase.
When the fluid is dust, $n_T=0$, when it is radiation, $n_T=2$.

The power spectrum is

\begin{equation}
\label{ps}
{\cal{P}} \propto n^3 |\mu_n|^2 \equiv (\Delta h_n)^2 \propto n^{n_T},
\end{equation}
where $\Delta h_n$ is the strain of the gravitational wave with wavenumber $n$.
Hence, one can obtain the spectra of the strain at the moment $\eta_c$ of curvature scale crossing
at dust domination and radiation domination from Eq.~(\ref{nt}), and they are
$\Delta h_n(\eta_c)\propto n^{0}$ and $\Delta h_n(\eta_c)\propto n$, respectively.

However, we need the strain today, $\Delta h_n(\eta_0)$. Note that
the curvature scale is given by $l_c\equiv (a^3/a'')^{1/2}$ (which is equal to the Hubble radius
in the case of a perfect fluid dominated classical model), and it is comparable to the physical wavelength
of the mode $k=n/R_H$, $l_{\rm phys} = a/k$, when $k^2=a''/a$. Hence, for $l_{\rm phys} < l_c$,
or $k^2>a''/a$, $v_k$ oscillates (see the mode equation (\ref{modeequation}) for $\Ka =0$).
Therefore, as $v_k =a \mu_k$,
one has that $|\mu_k|(\eta_0) a(\eta_0)=|\mu_k|(\eta_c) a(\eta_c)$ after curvature scale crossing
at the expanding phase. Note that $a(\eta_c)$ as a function of $k$ can be obtained through

\begin{equation}
l_{\rm phys}(\eta_c) = a(\eta_c)/k=l_c(\eta_c) \propto R_H(\eta_c) \propto t(\eta_c) \propto a^{3/2}(\eta_c)
\end{equation}
when $\eta_c$ is in the dust dominated era, and
\begin{equation}
l_{\rm phys}(\eta_c) = a(\eta_c)/k=l_c(\eta_c) \propto R_H(\eta_c) \propto t(\eta_c) \propto a^2(\eta_c)
\end{equation}
when $\eta_c$ is in the radiation dominated era. Hence one gets that
$a(\eta_c) \propto k^{-2}$ and $a(\eta_c)\propto k^{-1}$ for dust and radiation dominated
eras, respectively. Therefore, in terms of the frequency $f=k/(2 \pi a_0)=nH_0/(4 \pi)$, the transfer function for
the strain,

\begin{equation}
\label{a}
\Delta h_f(\eta_0)=T_f\Delta h_f(\eta_c),
\end{equation}
which is the same as the one for $|\mu_f|$,
see Eq.~(\ref{ps}), reads

\begin{equation}
\label{b}
T_f \propto \left(\frac{H_0}{f}\right)^2\left(1+\frac{f}{f_{\rm eq}}\right),
\end{equation}
where $f_{\rm eq}=H_{\rm eq}a_{\rm eq}/a_0 \approx 10^{-16}$ Hz and $H_0 \approx 10^{-18}$ Hz.

Finally, the potential $V(\bar{\eta})=(d^2a/d\bar{\eta}^2)/a$ of Eq.~(\ref{FinalModeEq}) derived from Eq.~(\ref{FinalScaleF}) reads

\begin{equation}
V(\bar{\eta})=\frac{2\left[\frac{1}{\bar{\eta}_*}+\frac{\bar{\eta}_b^2}{\left(\bar{\eta}_b^2+
\bar{\eta}^2\right)^{3/2}}\right]}{\frac{\bar{\eta}^2}{\bar{\eta}_*}+2\sqrt{\bar{\eta}_b^2+\bar{\eta}^2}},
\end{equation}
and it has a maximum value at $\bar{\eta}=0$, $V_{\rm
max}\approx\bar{\eta}_b^{-2}$. We see therefore that modes with
$n>1/\bar{\eta}_b$ do not feel the potential at any time, for they
correspond to wavelengths much smaller than the maximum curvature
scale of the model. Hence, they are always oscillating with
spectrum given by the initial vacuum state,

\begin{equation}
v_n(\eta)=\frac{e^{-in\bar{\eta}}}{\sqrt{2n}},
\end{equation}
and there is no graviton creation above these frequencies.
Therefore, we make a cutoff at the point $n>1/\bar{\eta}_b$, so
that there is no zero-point contributions to the energy density,
and the vacuum spectrum $\Omega_{GW}$ is clearly finite. Actually,
it is important to stress that such zero-point contribution would
affect only the high frequency sector of our results.

From Eq.~(\ref{ps}), one then infers that the strain $\Delta h_n$ has spectrum $n$ for $n>1/\bar{\eta}_b$,
independently of $\eta$. As the bounce itself has a duration much smaller then the classical dust-radiation
evolution, almost all modes that do not enter the curvature scale after the beginning of classical
radiation domination in the expanding phase have the above spectrum.

Noting that our model is symmetric with respect to the bounce, implying that wavelengths which leave the curvature
scale before the bounce at some fluid domination era enter the curvature scale after the bounce at the same
fluid domination, we can summarize the results concerning the spectrum of the strain in the following way:

i) $H_0\approx 10^{-18} {\rm Hz} < f < 10^{-16} {\rm Hz} \approx f_{\rm eq};\;\;\;$ \\
$\Delta h_f(\bar{\eta}_0)\propto f^{-2}f^0 = f^{-2}$.

ii) $f_{\rm eq}\approx 10^{-16} {\rm Hz} < f < f_{\rm max} = H_0/\bar{\eta}_b;\;\;\;$ \\
$\Delta h_f(\bar{\eta}_0)\propto f^{-1}f = f^{0}$.

iii) $f_{\rm max} = H_0/\bar{\eta}_b < f;\;\;\;$ $\Delta h_f(\bar{\eta}_0)\propto f$.

From the above spectrum, one can easily obtain the spectrum for the energy density
parameter of gravitational waves from the relation

\begin{equation}
\Omega_{GW}(f,\bar{\eta}_0) \propto f^2 [\Delta h_f(\bar{\eta}_0)]^2 .
\end{equation}

These results are confirmed by the numerical calculation we show in the sequel.

\subsection{Numerical Results} \label{sec:lambconst}
\label{subsec:numres}

\begin{figure}[htbp]
\begin{center}
\includegraphics[scale=0.95]{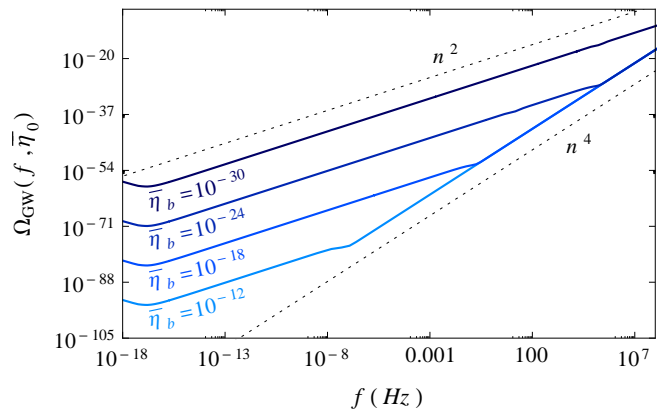}
\caption{Energy density parameter as a function of frequency for the primordial gravitational waves produced in our model. Each curve shows the results obtained for the indicated value of $\bar{\eta}_b$. The dashed lines show for comparison two power laws proportional to $n^2$ and $n^4$.}  \label{Omega}
\end{center}
\end{figure}

\begin{figure}[htbp]
\begin{center}
\includegraphics[scale=0.95]{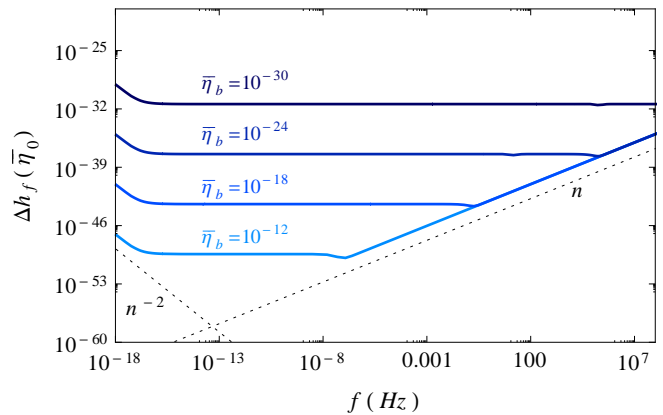}
\caption{Strain amplitude as a function of frequency for the same primordial gravitational waves produced in our model. Each curve shows the results obtained for the indicated value of $\bar{\eta}_b$. The dashed lines show for comparison two curves proportional to $n^{-2}$ and $n$.}  \label{Deltah}
\end{center}
\end{figure}

Fig. \ref{Omega} shows the value of the density parameter as a function of frequency $f$ for the four different values of $\bar{\eta}_b$ we are considering.
Fig. \ref{Deltah} shows the results of the strain amplitude
\begin{equation}
[\Delta h_n(\bar{\eta}_0)]^2= 6 \left( \frac{a_0H_0}{n} \right)^2 \Omega_{GW}(n,\bar{\eta}_0).
\end{equation}

Both figures are in accordance with our previous analytical considerations.

\begin{figure}[htbp]
\begin{center}
\includegraphics[scale=0.95]{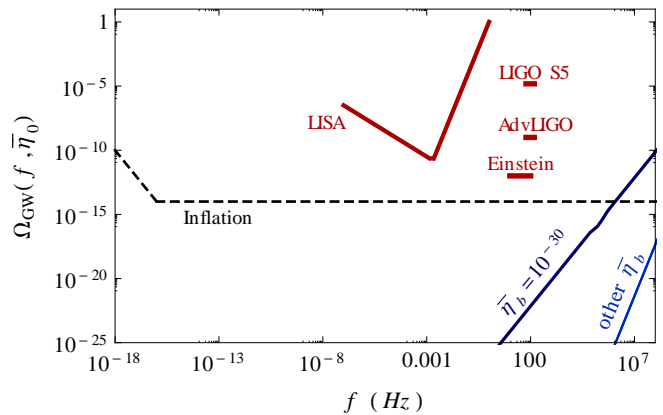}
\caption{Comparison of our results (blue curves at the bottom right) with experimental sensitivities (red curves) and a prediction the upper limits on the spectrum of primordial gravitational waves generated in inflationary models (black dashed curve). The red curves show the sensitivities achieved by LIGO's 5th run and the ones predicted for Advanced LIGO, the Einstein Telescope and LISA \cite{Detectors}. See \cite{Nature} and references therein.}  \label{CompExp}
\end{center}
\end{figure}

In order to establish the possibility of detecting the primordial gravitational waves produced in our theoretical model, we show in Fig. \ref{CompExp} a zoom of the upper part of Fig. \ref{Omega}, together with other theoretical predictions and experimental sensitivities.

We see that only for the lowest value of $\bar{\eta}_b$ we studied, $\bar{\eta}_b=10^{-30}$, and for a very high value of frequency, does our result reach the upper limits predicted by the inflationary model. Not even in this extreme case, is the stochastic background strong enough to be directly detected by present and future planned experiments. Nonetheless, we must emphasize that, while our calculations are exact, the predictions presented for inflation are only upper limits and could, therefore, be much lower than the curve presented in Fig. \ref{CompExp}.

\section{Conclusions}

In this paper we have calculated the amplitude and spectrum of energy density and strain of relic gravitons in a quantum bouncing cosmological model. The strain spectrum of this quantum bouncing model is different from the cyclic and inflationary
scenarios. While these two models have spectra $\approx k^{-2}, \approx k^{-1}$ at dust domination, and $\approx k^{-1}, \approx k^{0}$ at radiation domination, respectively, our model have spectra $\approx k^{-2}$ and $\approx k^{0}$ at the same eras.

One possible different scenario in which the power of relic gravitons could be enhanced could be obtained by adding to the matter content of the model some amount of stiff matter, which has primordial spectral index $n_T =2$. This will be the subject of future investigations.

As a final point, as in the cyclic ekpyrotic model, the resulting amplitude is too small to be detected by any gravitational wave detector. In particular, the sensitivity of the future third generation of gravitational wave detectors, as for example the Einstein Telescope, could reach $\Omega_{\rm GW} \sim 10^{-12}$ at the frequency range $10-100\,{\rm Hz}$ to an observation time of $\sim 5$ years and with a signal-to-noise ratio $({\rm S/N})\sim 3$. Therefore, any detection of relic gravitons, in this frequency range, will rule out this type of quantum bouncing model as a viable cosmological model of the primordial universe.

\begin{acknowledgments}
DB thanks the Brazilian agency FAPESP, grant 2009/15612-6 for financial support, and ICRA-CBPF for its kind hospitality. BBS would like to acknowledge the financial support of CAPES under grant number CAPES-PNPD 2940/2011 and to thank CNPq for the 2010-2011 PCI fellowship. NPN would like to thank CNPq for financial support. ODM would like to thank the Brazilian agency CNPq for partial financial support (grant 300713/2009-6). The authors also thank Sandro Vitenti for his kind help with numerical methods.

\end{acknowledgments}

\end{document}